\documentclass[journal]{IEEEtran}
\usepackage{xcolor}
\usepackage{stmaryrd}
\usepackage{amsmath}
\usepackage{amsthm}
\usepackage{xfrac}
\usepackage{amssymb}
\usepackage{mathabx}
\usepackage{psfrag}
\usepackage{graphicx}
\usepackage{epstopdf}
\usepackage{todonotes}
\usepackage{cite}
\usepackage{algorithmic}
\usepackage{algorithm}
\usepackage{svg}
\usepackage{mathtools}
\usepackage{hyperref}
\usepackage{hyphenat}
\usepackage{booktabs}
\usepackage{accents}
\usepackage{bbm}
\usepackage{comment}
\usepackage{balance}
\usepackage{mathtools,amssymb,lipsum}

\usepackage{cuted}
\newlength{\dhatheight}

\usepackage{float}
\usepackage{multirow}

\usepackage[caption=false,font=footnotesize]{subfig}

\usepackage{xparse,mathtools}

\begin{document}
%

\title{
Spreading over OFDM for Integrated Sensing and Communications (ISAC) Ranging: Multi-user Interference Mitigation
}

\author{Karim A. Said, Ibraheem Alturki, Husheng Li, Elizabeth Bentley, and Lingjia Liu
\thanks{K. Said, Ibraheem Alturki and L. Liu are with Wireless@Virginia Tech, the Bradley Department of ECE at Virginia Tech, Blacksburg, VA. Husheng Li is with the School of Aeronautics and Astronautics, Purdue University, PU. E. Bentley is with the Information Directorate of Air Force Research Laboratory, Rome NY. 
The corresponding author is L. Liu (ljliu@ieee.org).}
\thanks{Distribution A. Approved for public release: Distribution Unlimited: AFRL-2025-2033 on 22 Apr 2025.}
}
%




\maketitle
\begin{abstract}
In the context of communication-centric integrated sensing and communication (ISAC), the orthogonal frequency division multiplexing (OFDM) waveform was proven to be optimal in minimizing ranging sidelobes when random signaling is used. A typical assumption in OFDM-based ranging is that the max target delay is less than the cyclic prefix (CP) length, which is equivalent to performing a \textit{periodic} correlation between the signal reflected from the target and the transmitted signal. In the multi-user case, such as in Orthogonal Frequency Division Multiple Access (OFDMA), users are assigned disjoint subsets of subcarriers which eliminates mutual interference between the communication channels of the different users. However, ranging involves an aperiodic correlation operation for target ranges with delays greater than the CP length. Aperiodic correlation between signals from disjoint frequency bands will not be zero, resulting in mutual interference between different user bands. We refer to this as \textit{inter-band} (IB) cross-correlation interference. In this work, we analytically characterize IB interference and quantify its impact on the integrated sidelobe levels (ISL). We introduce an orthogonal spreading layer on top of OFDM that can reduce IB interference resulting in ISL levels significantly lower than for OFDM without spreading in the multi-user setup. We validate our claims through simulations, and using an upper bound on IB energy which we show that it can be minimized using our proposed spreading. However, for orthogonal spreading to be effective, a price must be paid in terms of spectral utilization, which is yet another manifestation of the trade-off between sensing accuracy and data communication capacity.

\end{abstract}


\section{Introduction}
Integrated Sensing and Communication (ISAC) has emerged as a key enabler in 6G networks, offering the potential to unify wireless communication and environmental sensing using a common set of spectral and hardware resources. This convergence addresses the growing demand for situational awareness  in applications such as autonomous systems, smart manufacturing, and next-generation mobility. ISAC is considered a fundamental component of 6G due to its ability to enhance spectral efficiency, reduce hardware redundancy, and enable new services \cite{liu2022integrated, 10349862}. Many 6G scenarios are expected to rely heavily on sensing for environment awareness \cite{9830717,10012421}, as well as for assisting the communication link \cite{10214383}.
Combining sensing and communication functions on the same hardware can enable size and weight reduction, which can be beneficial for UAV-based platforms \cite{9050413}.

A common theme in ISAC is the conflict between the requirements of communication and sensing functions. One example is the tradeoff between communication performance in terms of capacity, and radar performance in terms of sidelobe level \cite{sahin2017characterization}. As ISAC systems evolve into multi-user ISAC, a new dimension is added to the sensing-communication interaction due to resource sharing between multiple users. Thus, \textit{inter-user} sensing-communication interactions arise, in addition to the \textit{intra-user} sensing communication interaction for the single user case.

Recent studies have explored different strategies to optimize multi-user ISAC systems. For instance, some works address multi-user interference management using Rate-Splitting Multiple Access (RSMA), a promising method to enhance the system's performance in handling both communication and sensing tasks simultaneously \cite{chen2024rate}. Other approaches explore joint beamforming strategies in multi-user MIMO setups that manage interference effectively, enabling radar and communication to coexist with minimal performance degradation \cite{liu2018mu}. Transceiver optimization in mmWave and THz multi-user ISAC systems has also been proposed to further mitigate interference and optimize system performance, especially in high-frequency environments \cite{wang2024joint}. Several works also focus on optimizing beamforming for multi-user and multi-target ISAC systems, addressing the trade-offs between sensing and communication \cite{zhu2023information}. Additionally, Reconfigurable Intelligent Surface (RIS)-assisted ISAC systems have been investigated, offering efficient interference management in multicell environments by adjusting the RIS phase shifts and beamforming strategies \cite{yang2024ris}. These advancements demonstrate the growing interest in multi-user ISAC system enhancement.


From a communication-centric ISAC standpoint, the ranging performance of OFDM waveforms has been studied in several works.  This is driven by the fact that OFDM is ubiquitous in a wide base of existing wireless devices \cite{8412469}, and has robust performance that requires equalization of low complexity. In addition,
the MIMO framework aligns well with the OFDM signaling structure. For sensing, the time-frequency
structure of a OFDM frame can map easily to a 2D range-speed radar information representation. The
subcarrier dimension, known as fast time, is used to measure the delay of objects in the environment, while the symbol dimension,  known as slow time, captures changes as a result of Doppler
due to the mobility of objects in the environment \cite{10770016,9529026}. In \cite{liu2024ofdm}, a generic waveform framework was used to evaluate the ranging performance for arbitrary communication waveforms in terms of the integrated sidelobe level (ISL). OFDM was proven to be globally optimal in minimizing the \textit{periodic} auto-correlation sidelobe energy when \textit{sub-Gaussian} constellations are used  i.e., constellations with kurtosis values less than $2$. Recognizing the optimality of OFDM, techniques to further explore the data rate ISL trade-off space do not involve changing the waveform but only manipulating the subcarrier modulation data.   In \cite{9839260}, data rate in terms of modulation order vs. ISL has been considered demonstrating the superiority of QPSK to higher order modulations. A more comprehensive analysis of the trade-off between general modulation constellation and ISL is provided in \cite{du2023reshaping}.  For metrics other than ISL, power allocation across subcarriers is used as a design variable to maximize capacity under the constraint of maintaining a given Peak sidelobe Level (PSL) level \cite{9337375}. For MIMO systems, exploiting both the spatial and temporal degrees of freedom, a method known as symbol level precoding (SLP) was used to reduce waveform sidelobes \cite{10622269}.

The global optimality of the OFDM waveform established in \cite{liu2024ofdm} is for the periodic auto-correlation case, but not for aperiodic auto-correlation. We revisit the problem of selecting the optimal waveform when target delays are longer than CP length, i.e., for the aperiodic correlation case. In addition, we consider the multi-user case where different groups of subcarriers are assigned to different users to transmit their signals. For monostatic ISAC, each user will be receiving the back-scatttered signal over its corresponding frequency subcarrier range that was used for transmission. Aperiodic correlation used for the ranging function will result in mutual inter-band interference (IB) from the transmissions of other users. Although the leakage is typically small, IB can be significant compared to the power of the in-band back-scattered in-band signal. In this work, we analyze IB interference in an OFDMA ISAC setting. We provide analytical expressions to quantify IB and its effect on ISL performance. Furthermore, we introduce a spreading layer that can minimize IB interference at the cost of a small loss in capacity. In a future work, we will further pursue the investigation of the optimality of our introduced spreading layer.
The following is a list of the main contributions of this work:
\begin{itemize}
\item We provide an analytical framework for representing the aperiodic correlation as a periodic correlation, thus enabling manipulation in the frequency domain.
\item In Proposition 1, we derive expressions for the normalized average square correlation function (ACF) for the two-user ISAC case when the OFDM signal used for target illumination consists of $M$ OFDM consecutive symbols.
\item In Corollary 1, we derive the analytical expression for the normalized average integrated sidelobe energy (EISL) for the 2 user ISAC case and $M$ OFDM symbols and identify the contribution of the IB interference energy $E_{IB}$.
\item In Theorem 1, we derive an upper bound on $E_{IB}$. Consequently, we derive a spreading method that can minimize the upper bound while maintaining the highest spectral utilization.
\end{itemize}

The remainder of the paper is organized as follows: Section II provides the system model and the mathematical framework for representing aperiodic correlation in the frequency domain. In Section III, two user Multi-User ISAC (MU-ISAC) scenario is considered, and closed form expressions for ACF and EISL are derived. In Section IV, an OFDM spreading layer based on periodic discrete prolate spheroidal sequences (P-DPSS) is presented which can minimize the out-of-band correlation interference in exchange for a minimal loss in spectral efficiency. Section V contains simulation results for validating the theoretical results of Proposition 1, Corollary 1 and Theorem 1, comparing OFDMA with and without the proposed spreading. Conclusions and directions for future extensions are provided in section VI.

\begin{figure}
\centering 
\includegraphics[width=1\linewidth]{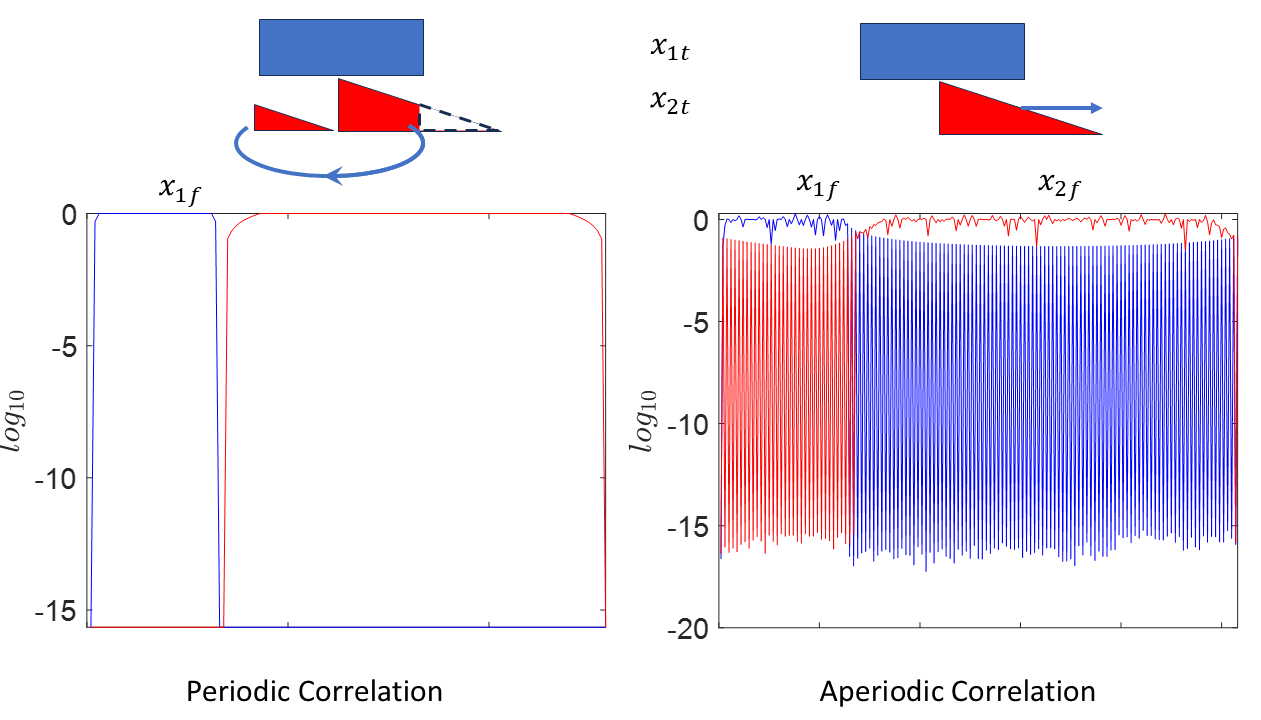}
\caption{(Left) Periodic/circular convolution. (Right) Aperiodic/linear convolution.} \label{adj_f_corr_lkg}
\end{figure}

\section{System Model}
We start with a OFDM signal based on a general waveform comprised of an orthonormal basis $\mathbf{O} \in \mathbb{C}^{N\times N}$, where each orthogonal component waveform is modulated by an information symbol vector $\mathbf{s} \in \mathcal{A}^{N\times 1}$ resulting in a discrete time signal in discrete time by $\mathbf{x}_t \in \mathbb{C}^{N \times 1}$ \eqref{OFDM_mod} 
\begin{equation}\label{OFDM_mod}
\mathbf{x}_t = \mathbf{O}\mathbf{s}
\end{equation}
For OFDM, $\mathbf{O}=\mathbf{F}$ where $\lbrace\mathbf{F}\rbrace_{l,k}=\frac{1}{\sqrt{N}}e^{-j2\pi \frac{lk}{N}}, l,k=0,..,N-1$.

In a monostatic ISAC, the information carrying signal $\mathbf{x}_t$ is transmitted by a given node, and the returned signal $\mathbf{y}_t$ consisting of reflections by targets is received by the same node and correlated against the original transmitted signal according to \eqref{corr_aprdc} \cite{9839260,9771849}

\begin{equation}\label{corr_aprdc}
c_{\mathbf{x}\mathbf{y}}[k] = \mathbf{x}_t^H \mathbf{J}_k \mathbf{y}_t ,\quad \mathbf{J}_k = \begin{bmatrix} \mathbf{0} & \mathbf{I}_{N-k} \\ \mathbf{0} & \mathbf{0} \end{bmatrix} \quad k=0,..,N-1 
\end{equation}
In the presence of a cyclic prefix, for a correlation lag range less than the CP length $l_{cp}$, \textit{periodic} correlation applies

\begin{equation}\label{corr_prdc}
\tilde{c}_{\mathbf{x}\mathbf{y}}[k] = \mathbf{x}_t^H \tilde{\mathbf{J}}_k \mathbf{y}_t, \quad \tilde{\mathbf{J}}_k = \begin{bmatrix} \mathbf{0} & \mathbf{I}_{N-k} \\ \mathbf{I}_k & \mathbf{0} \end{bmatrix} \quad k=0,..,l_{cp} 
\end{equation}
where  

\begin{equation} \label{corr_circ}
\tilde{\mathbf{J}}_k =  \mathbf{F}_{N}^H D(\mathbf{z}_N^{\frac{k}{N}}) \mathbf{F}_{N} \end{equation}
where $\mathbf{z}_N=[z^0,..,z^{N-1}]$,  $z=e^{j\pi}$, and $\textit{D}(\mathbf{a})$ is a diagonal matrix with $\mathbf{a}$ on the diagonal. We distinguish between aperiodic correlation $c_{\mathbf{x}\mathbf{y}}$, and periodic correlation $\tilde{c}_{\mathbf{x}\mathbf{y}}$ using $\tilde{ }$ for the latter.
 
Using \eqref{corr_circ}, it is possible to express the periodic correlation in the frequency domain as a point-wise multiplicative operation in \eqref{corr_fdomain}
\begin{equation}\label{corr_fdomain}
\tilde{\mathbf{c}}_{\mathbf{x}\mathbf{y}} = \mathbf{F}_N^H(\mathbf{x}_f^*\odot \mathbf{y}_f)
\end{equation}
where $\mathbf{x}_f = \mathbf{F}_N\mathbf{x}_t$, $\mathbf{y}_f = \mathbf{F}_N\mathbf{y}_t$, and$\mathbf{c}_{\mathbf{x}\mathbf{y}} =[c_{\mathbf{x}\mathbf{y}}[0],..c_{\mathbf{x}\mathbf{y}}[N],c_{\mathbf{x}\mathbf{y}}[-(N-1)],..,c_{\mathbf{x}\mathbf{y}}[-1]]$.

\subsection{Aperiodic correlation in the frequency domain}
By applying zero-padding, it is possible to represent the aperiodic correlation as a multiplicative operation in the frequency domain. 

Let $\bar{\mathbf{x}}_t = [\mathbf{x}_t^T,\mathbf{0}_{zp}^T]^T, \bar{\mathbf{y}}_t = [\mathbf{y}_t^T,\mathbf{0}_{zp}^T]^T$ , where $zp > N-1$ we have the following identity
\begin{equation}\label{prdc_aprdc_eq}
\begin{split}
\mathbf{c}_{\mathbf{x}\mathbf{y}}[k]&=\tilde{\mathbf{c}}_{\bar{\mathbf{x}}\bar{\mathbf{y}}}[k]\\
&=\bar{\mathbf{x}}_t^H\mathbf{F}_{M}^HD(\mathbf{z}_M^{\frac{k}{M}})\mathbf{F}_{M}\bar{\mathbf{y}}_t\\
&=\bar{\mathbf{x}}_f^HD(\mathbf{z}_M^{\frac{k}{M}}) \bar{\mathbf{y}}_f\\
\end{split}
\end{equation}
where $M=N+zp$. 
\begin{figure}
\centering 
\includegraphics[width=1\linewidth]{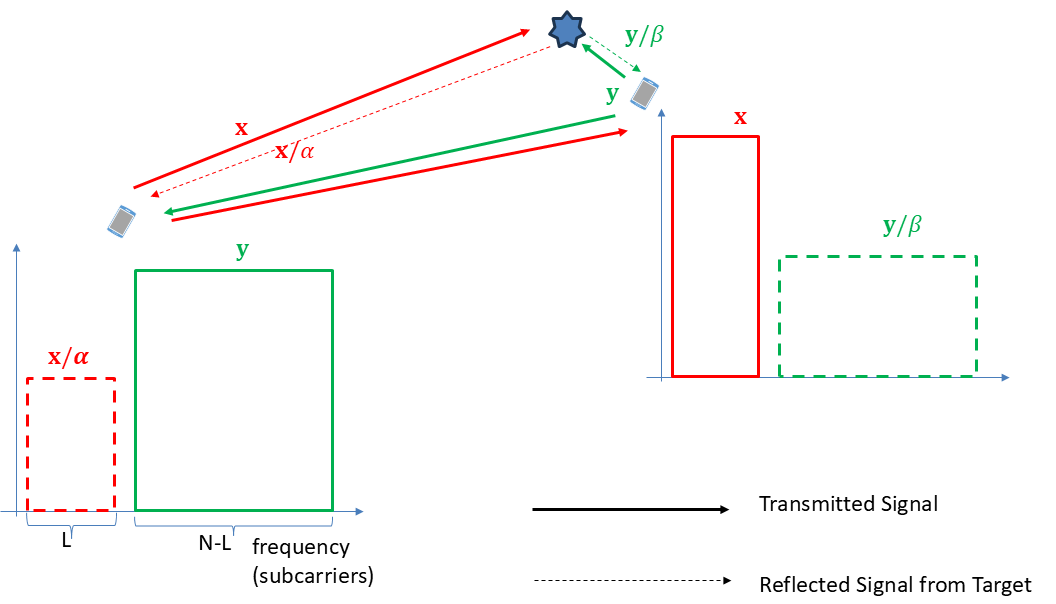}
\caption{Two-user OFDMA-ISAC} \label{OFDMA_ISAC}
\end{figure}

Setting $zp=N$. the double-sided correlation can be written in vector form
\begin{equation}\label{f_corr}
\mathbf{c}_{\mathbf{x}\mathbf{y}} = \mathbf{F}_{2N}^H \left(\bar{\mathbf{x}}_f^*\odot \bar{\mathbf{y}}_f\right)
\end{equation}
Equation \eqref{f_corr} can be visualized as the product of the red and blue curves in the graph on the right of Fig. \ref{adj_f_corr_lkg}. Periodic correlation is shown on the left of Fig. \ref{adj_f_corr_lkg}, which shows that two disjoint signals in the frequency domain will result in zero correlation. However, due to the time-domain zero-padding involved in the aperiodic correlation given by \eqref{f_corr}, an upsampling-by-2 (for $zp=N$) operation occurs in the frequency domain which causes mutual leakage between two non-overlapping bands and non-zero correlation.

We consider a frequency band consisting of $N$ discrete frequencies divided between two users, user-1 and user-2. A band consisting of $L$ frequencies allocated to user-1 and a band of $N-L$ frequencies allocated to user-2. This is depicted in the bottom left  part of Fig. \ref{adj_f_corr_lkg}, blue for user-1 and red for user-2. 
Starting from \eqref{OFDM_mod}, we develop the following relations between the modulation symbol vectors $\mathbf{s}^{(1)},\mathbf{s}^{(2)}$ and their corresponding frequency domain quantities $\bar{\mathbf{x}}^{(1)}_{f}, \bar{\mathbf{x}}^{(2)}_{f}$, for user-1 and user-2, in \eqref{leaky_x1f} and \eqref{leaky_x2f}, respectively. $\mathbf{P}^{(1)}\in \mathbb{C}^{L\times L}, \mathbf{P}^{(2)} \in \mathbb{C}^{(N-L)\times (N-L)}$ represent frequency domain spreading matrices applied to the symbols of user-1 and user-2, respectively.

\begin{equation}\label{leaky_x1f}
\begin{split}
\mathbf{\bar{x}}^{(1)}_{f}
& = \mathbf{F}_{2N} \mathbf{\bar{x}}^{(1)}_{t}\\
&= \mathbf{F}_{2N} \begin{bmatrix} \mathbf{F}_{N} & \mathbf{0}_N \end{bmatrix}^H \begin{bmatrix} \mathbf{I}_L & \mathbf{0}_{N-L} \end{bmatrix} ^H\mathbf{P}^{(1)}\mathbf{s}^{(1)}\\
&=\textit{D}\left(\mathbf{z}_{2N}^{\frac{N-1}{2N}}\right) \mathbf{B}^{(1)}\mathbf{U}_L\textit{D}\left(\mathbf{z}_{L}^{\frac{N-1}{2N}}\right)\mathbf{P}^{(1)}\mathbf{s}^{(1)}\\
&= \mathbf{W}^{(1)}\mathbf{s}^{(1)}
\end{split}
\end{equation}
\begin{equation}\label{leaky_x2f}
\begin{split}
\mathbf{\bar{x}}_{f}^{(2)}
& = \mathbf{F}_{2N} \mathbf{\bar{x}}_{t}^{(2)}\\
&= \mathbf{F}_{2N} \begin{bmatrix} \mathbf{F}_{N} & \mathbf{0}_N \end{bmatrix}^H \begin{bmatrix} \mathbf{0}_L & \mathbf{I}_{N-L} \end{bmatrix} ^H\mathbf{P}^{(2)}\mathbf{s}^{(2)}\\
&=\textit{D}\left(\mathbf{z}_{2N}^{\frac{N-1}{2N}}\right) \mathbf{B}^{(2)}\mathbf{U}_{N-L}\textit{D}\left(\mathbf{z}_{N-L}^{\frac{N-1}{2N}}\right)\mathbf{P}^{(2)}\mathbf{s}^{(2)}\\
&= \mathbf{W}^{(2)}\mathbf{s}^{(2)}
\end{split}
\end{equation}
$\mathbf{U}_L=[\mathbf{e}_1,\mathbf{0}, \mathbf{e}_2,..,\mathbf{0},\mathbf{e}_L]^T$, i.e, up-sampling by 2, where $[\mathbf{e}_i]_k=\delta(k-i), k =0,..,L-1 $. 
$[\mathbf{B}]_{k,k'}=\frac{\sin(N(k - k')\frac{\pi}{2N})}{\sin((k - k')\frac{\pi}{2N})}$ $ \text{ for }k,k'=0,..,2N-1$,
\begin{equation}
\mathbf{B}= \begin{bmatrix}
\mathbf{B}^{(1)} & \mathbf{b}_{2L-1} &\mathbf{B}^{(2)} & \mathbf{b}_{2N-1} \\
\end{bmatrix}\\
\end{equation}
where $\mathbf{b}_r=\frac{\sin(N(k - r)\frac{\pi}{2N})}{\sin((k - r)\frac{\pi}{2N})}$   $ \text{ for }k=0,..,2N-1$.

All factors in the third line of each of  for \eqref{leaky_x1f} and \eqref{leaky_x2f} are lumped into $\mathbf{W}^{(1)} \in \mathbb{C}^{2N\times L}, \mathbf{W}^{(2)} \in \mathbb{C}^{2N\times (N-L)}$, respectively. The derivations of the fourth line in each of \eqref{leaky_x1f} and \eqref{leaky_x2f} are provided in Appendix \ref{apndx_A}.  
.
\section{Quantifying Leakage Energy in Two user OFDMA ISAC}

\subsection{ISAC ranging in two-user OFDMA scenario}
Consider the scenario shown in Fig. \ref{OFDMA_ISAC} where user-1 is allocated the range of subcarriers depicted by the red color and user-2 is allocated subcarriers depicted by the green color. Without loss of generality, user 1 illuminates a target by signal $\mathbf{x}_{t1}$ and the reflected signal is attenuated by a factor $\alpha$ and received by
user 1 to perform the ranging function. A transmitted signal from user-2 can reach user 1 through either a line-of-sight (LOS) path or non-LOS (NLOS) paths from objects in the environment. Assuming an LOS path from user-1 to user-2 exists, the effect of the correlation leakage can have a significant impact on the ranging function. In what follows, we analyze the correlation side-lobes in the presence of correlation leakage due to an adjacent frequency signal source. 

Assuming the discrete time domain received signal at user-1 is given by $\mathbf{y}_t=\mathbf{x}_{t}^{(1)}+\alpha\mathbf{x}_{t}^{(2)}$, we find the  $k$-th element of the correlation vector using \eqref{f_corr} and thus obtain 
\begin{equation}\label{2user_corr}
\mathbf{c}_{xy}[k] =\mathbf{F}_{2N}^H[k]\left(  \left(\bar{\mathbf{x}}_{f}^{(1)*}\odot \bar{\mathbf{x}}_{f}^{(1)}\right)  +   \alpha\bar{\mathbf{x}}_{f}^{(1)*}\odot \bar{\mathbf{x}}_{f}^{(2)}\right)
\end{equation}
The first term in \eqref{2user_corr} corresponds to the auto-correlation of user 1's transmitted signal, and the second term corresponds to adjacent frequency correlation interference from user 2 to user 1. We note that the main correlation lobe energy is unaffected by the adjacent frequency interference and remains the same as in the single user case $\mathbf{c}_{xy}[0]=||\mathbf{s}||_2^4$.

\subsection{Correlation using Frame of $M$ OFDM symbols}

\begin{figure}
\centering 
\includegraphics[width=1\linewidth]{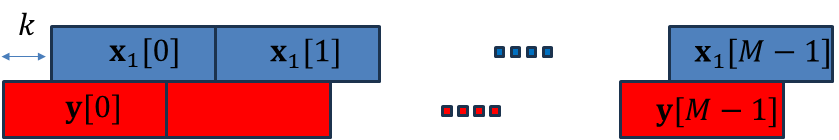}
\caption{Correlation of a block of $M$ symbols} \label{Blk_Msym_corr}
\end{figure}

We consider in this section the case when a signal comprised of a sequence of $M$ consecutive OFDM symbols is used to illuminate a target. The vector of the incident signal on the target becomes $\mathbf{x}_{\mathbf{M}}\triangleq [\mathbf{x}^{(1)T}[0],..,\mathbf{x}^{(1)T}[M-1]]^T$, and the backscattered signal vector is $\mathbf{y}_{\mathbf{M}}\triangleq [\mathbf{y}[0]^T,..,\mathbf{y}^T[M-1]]^T$ where $\mathbf{y}[i]=\mathbf{x}^{(1)}[i]+\mathbf{x}^{(1)}[i]$. As illustrated  in Fig. \ref{Blk_Msym_corr}, the correlation based on a block of $M$ symbols is composed of the sum of correlations between incident and reflected symbols having the same index, and correlations between incident and reflected symbols having indices that differ by $1$.

\begin{equation}\label{mult_sym_corr}
\begin{split}
&c_{\mathbf{x}_{\mathbf{M}},\mathbf{y}_{\mathbf{M}}}[k]\\
&=\sum_{i=0}^{M-1} c_{\mathbf{x}_1[i],\mathbf{y}[i]}[k] + \sum_{i=0}^{M-2} c_{\mathbf{x}^{(1)}[i],\mathbf{y}[i+1]}[-(N-k)]\\
&=\sum_{i}^{M-1} \bar{\mathbf{x}}_{f}^{(1)H}[i]D(\mathbf{z}_{2N}^{\frac{k}{2N}})\bar{\mathbf{y}}_f[i]\\
&+ \sum_{j=0}^{M-2} \bar{\mathbf{x}}_{f}^{(1)H}[j]D(\mathbf{z}_{2N}^{\frac{-(N-k)}{2N}})\bar{\mathbf{y}}_f[j+1]\\
&=\sum_{i}^{M-1} \bar{\mathbf{x}}_{f}^{(1)H}[i]D(\mathbf{z}_{2N}^{\frac{k}{2N}})(\bar{\mathbf{x}}_{1f}[i]+\alpha \bar{\mathbf{x}}_{f}^{(2)}[i])  \\&+\sum_{j=0}^{M-2}\bar{\mathbf{x}}_{f}^{(1)H}[j]D(\mathbf{z}_{2N}^{\frac{-(N-k)}{2N}})(\bar{\mathbf{x}}_{f}^{(1)}[j+1]+\alpha \bar{\mathbf{x}}_{f}^{(2)}[j+1])\\ &\quad \quad\text{ for }  0 \leq k\leq N-1\\
\end{split}
\end{equation}
In the following proposition, we quantify the effects of the type of modulation of the in-band user, $\mathbf{s}^{(1)}$, and the out-of-band user, $\mathbf{s}^{(2)}$, and the spreading applied to each on the aperiodic autocorrelation function (A-ACF) from the point of view of user-1.

\newtheorem{Theorem}{Theorem}
\newtheorem{Proposition}{Proposition}

\newtheorem{Corollary}{Corollary}
\begin{Proposition}[Two user ISAC A-ACF magnitude squared, $M$ consecutive symbols]\label{Prop1}

The nomalized average squared A-ACF for an illuminating signal generated by user-1 comprised of $M$ consecutive OFDM symbols, in each OFDM symbol, $L$ subcarriers assigned to user-1, and $N-L$ assigned to other users.

\begin{equation}\label{corr_vec_sq_E}
\begin{split}
&\mathbb{E}(|\mathbf{c}_{\mathbf{x}_{\mathbf{M}},\mathbf{y}_{\mathbf{M}}}[k]|^2)\\
&=\frac{M(\mu_4-2)}{M^2L^2}\sum_{m,n}^{2N-1}||\mathbf{v}_n^{(1)} \odot \mathbf{v}_m^{(1)}||_2^2z^{\frac{k(n-m)}{N}}\\
&+\frac{M^2}{M^2L^2}\sum_{m,n}^{2N-1}||\mathbf{v}^{(1)}_n||^2 ||\mathbf{v}^{(1)}_m||^2z^{\frac{k(n-m)}{N}}\\
&+\frac{M}{(M^2L^2)}\sum_{m,n}^{2N-1}|\mathbf{v}^{(1)H}_n\mathbf{v}^{(1)}_m|^2z^{\frac{k(n-m)}{N}}\\
&+\frac{(M-1)}{(M^2L^2)}\sum_{m,n}^{2N-1}|\mathbf{v}^{(1)H}_n\mathbf{v}^{(1)}_m|^2(-z^{\frac{k}{N}})^{n-m}\\
&+\frac{\alpha^2M}{M^2L^2}\sum_{m,n}^{2N-1}\mathbf{v}^{(2)H}_m\mathbf{v}^{(2)}_n\mathbf{v}^{(1)H}_n\mathbf{v}^{(1)}_mz^{\frac{k(n-m)}{N}}\\
&+\alpha^2\frac{(M-1)}{M^2L^2}\sum_{m,n}^{2N-1}\mathbf{v}^{(2)H}_m\mathbf{v}^{(2)}_n\mathbf{v}^{(1)H}_n\mathbf{v}^{(1)}_m(-z^{\frac{k}{N}})^{n-m}\\
&\quad \quad\text{ for }  -(N-1) \leq k\leq N-1\\
\end{split}
\end{equation}

\end{Proposition}
where $\mathbf{V}^{(1)}=[\mathbf{v}^{(1)}_0,..,\mathbf{v}^{(1)}_{2N-1}]=\mathbf{W}^{(1)H}$,  $\mathbf{V}^{(2)}= [\mathbf{v}^{(2)}_0,..,\mathbf{v}^{(2)}_{2N-1}]=\mathbf{W}^{(2)H}$, and $\mathbf{W}^{(1)},\mathbf{W}^{(2)}$ are defined in \eqref{leaky_x1f} and \eqref{leaky_x2f}
, respectively.

\textit{Proof}. 
Appendix \ref{apndx_B}

We can see that in terms of modulation type, the dependence is only on the modulation type for user-1 through the kurtosis parameter $\mu_4$ in the second line. As for spreading, both user-1 and user-2 spreading have an impact.
The last two terms in \eqref{corr_vec_sq_E} represent the IB leakage. We note that for a very larger number of OFDM symbols in the illuminating signal, i.e.,  $M\rightarrow \infty$, the A-ACF reduces to \eqref{corr_vec_sq_E_M_infty}, showing that the IB effect disappears. However, $M$ cannot be too large in order to ensure a certain level of performance to track moving~targets.

\begin{equation}\label{corr_vec_sq_E_M_infty}
\begin{split}
&\mathbb{E}(|\mathbf{c}_{\mathbf{x}_{\boldsymbol{\infty}},\mathbf{y}_{\boldsymbol{\infty}}}[k]|^2)=\frac{1}{L^2}\sum_{m,n}^{2N-1}||\mathbf{v}^{(1)}_n||^2 ||\mathbf{v}^{(1)}_m||^2z^{\frac{k(n-m)}{N}}\\
\end{split}
\end{equation}
Equation \eqref{corr_vec_sq_E} can be written in matrix for as

\begin{equation}\label{corr_vec_sq_E_matrix}
\begin{split}
&\mathbb{E}(|\mathbf{c}_{\mathbf{x}_{\mathbf{M}},\mathbf{y}_{\mathbf{M}}}[k]|^2)\\
&=\frac{M(\mu_4-2)}{M^2L^2}\text{tr}\left((\mathbf{W}^{(1)H}\odot\mathbf{W}^{(1)T})\mathbf{z}\mathbf{z}^H(\mathbf{W}^{(1)H}\odot\mathbf{W}^{(1)T})^H\right)\\
&+\frac{M^2}{M^2L^2}\text{diag}(\mathbf{W}^{(1)}\mathbf{W}^{(1)H})^H\mathbf{z}_k\mathbf{z}_k^H\text{diag}(\mathbf{W}^{(1)}\mathbf{W}^{(1)H})\\
&+\frac{M}{(M^2L^2)}\mathbf{1}^H (\mathbf{W}^{(1)}\mathbf{W}^{(1)H}\odot\mathbf{W}^{(1)*}\mathbf{W}^{(1)T}\odot\mathbf{z}_k\mathbf{z}_k^H)\mathbf{1} \\
&+\frac{(M-1)}{(M^2L^2)}\mathbf{1}^H (\mathbf{W}^{(1)}\mathbf{W}^{(1)H}\odot\mathbf{W}^{(1)*}\mathbf{W}^{(1)T}\odot\tilde{\mathbf{z}}_k\tilde{\mathbf{z}}_k^H)\mathbf{1} \\
&+\frac{\alpha^2M}{M^2L^2}\mathbf{1}^H (\mathbf{W}^{(1)}\mathbf{W}^{(1)H}\odot\mathbf{W}^{(2)*}\mathbf{W}^{(2)T}\odot\mathbf{z}_k\mathbf{z}_k^H)\mathbf{1} \\
&+\alpha^2\frac{(M-1)}{M^2L^2}\mathbf{1}^H (\mathbf{W}^{(1)}\mathbf{W}^{(1)H}\odot\mathbf{W}^{(2)*}\mathbf{W}^{(2)T}\odot\tilde{\mathbf{z}}_k\tilde{\mathbf{z}}_k^H)\mathbf{1} \\
&\quad \quad\text{ for }  -(N-1) \leq k\leq N-1\\
\end{split}
\end{equation}




Further manipulation of the result of Proposition \ref{Prop1} gives us the expected value of the integrated side-lobe energy, EISL, in Corollary \ref{Corr1}.

\begin{Corollary}[Two user ISAC EISL]\label{Corr1}
The expected value of the integrated side-lobe energy (EISL) for the two user case is given by
\begin{equation}\label{EISL_final}
\begin{split}
&\text{EISL}\\
&= \frac{2N}{(ML)^2} \sum_{n=0}^{2N-1} ( M^2+2M-1)||\mathbf{v}^{(1)}_n||_2^4+M(\mu_4-2)||\mathbf{v}^{(1)}_n||_4^4\\
&+\frac{2N}{(ML)^2}E_{IB}- \frac{\mu_4 -1+ML}{(ML)^2}\\
\end{split}
\end{equation}
where

\begin{equation}\label{E_OCL}
\begin{split}
E_{IB}&=\alpha^2\sum_{n=0}^{2N-1}(2M-1)||\mathbf{v}^{(1)}_n||_2^2||\mathbf{v}^{(2)}_n||_2^2\\
&= \alpha^2(2M-1)\text{tr}\left(\mathbf{W}^{(1)}\mathbf{W}^{(1)H}\odot \mathbf{W}^{(2)}\mathbf{W}^{(2)H}\right)
\end{split}
\end{equation}
where $tr()$ denotes the trace operation. $E_{IB}$ represents  the inter-band cross-correlation energy contribution.
\end{Corollary}
\textit{Proof}. 
Appendix \ref{apndx_B}

\section{Effect of spreading on Out-of-band Correlation Leakage}

For the single user scenario, it was proven in \cite{liu2024ofdm} that OFDM is optimal in minimizing the EISL (for the periodic correlation operation). Therefore, spreading the energy of a single symbol on a distribution of sub-carriers is of no benefit. However, from \eqref{EISL_final} we can see that in the multi-user case a new factor is introduced, which is the out-of-band correlation leakage $E_{IB}$. Distributing the symbol energy across multiple sub-carriers using an orthonormal transform, $\mathbf{P}$ in \eqref{leaky_x1f}, in is equivalent to changing the modulation waveform.
Hence, the EISL minimizing waveform problem in the multi-user case has a different solution other than OFDM. The optimal solution must be one that minimizes not only the integrated side-lobe from correlation with the in-band signal, i.e., the back-scattered version of the illuminating signal, but also the contributions from out-band users, i.e., $E_{IB}$.  This minimization will depend on both the spreading used for the in-band user $\mathbf{P}^{(1)}$ and the out-of-band user. $\mathbf{P^{(2)}}$ In what follows, we analyze the dependence of $E_{IB}$ on the spreading of each user and derive an upper bound.

\begin{Theorem}[Upper bound on $E_{IB}$]\label{Prop2}$E$
For an in-band user signal spreading matrix $\mathbf{P}^{(1)}\in \mathbb{C}^{L\times L}$ and out-of-band user spreading matrix $\mathbf{P}^{(2)}\in \mathbb{C}^{(N-L)\times (N-L)}$, we have the following upper bound

\begin{equation}\label{E_OCL_Ubound}
E_{IB} \leq ||\bar{\mathbf{W}}^{(1)}||^2_F ||\tilde{\mathbf{W}}^{(2)}||^2_F + ||\tilde{\mathbf{W}}^{(1)}||^2_F ||\bar{\mathbf{W}}^{(2)}||^2_F 
\end{equation}

where $\mathbf{W}^{(1)}=\begin{bmatrix}
\bar{\mathbf{W}}^{(1)}\\
\tilde{\mathbf{W}}^{(1)}
\end{bmatrix}, \quad  \mathbf{W}^{(2)}=\begin{bmatrix}
\bar{\mathbf{W}}^{(2)}\\
\tilde{\mathbf{W}}^{(2)}
\end{bmatrix}$. 
\end{Theorem}
Appendix \ref{apndx_D}

Recalling the definitions of $\mathbf{W}^{(1)}$ and $\mathbf{W}^{(2)}$ given by \eqref{leaky_x1f} and \eqref{leaky_x2f}, since $\mathbf{P}^{(1)}, \mathbf{P}^{(2)}$ are orthonormal matrices, their specific choices will have no impact on the upper bound \eqref{E_OCL_Ubound}. However, with further examination of the other factors, in particular the matrices $\mathbf{B}^{(1)}, \mathbf{B}^{(2)}$ which are Dirichilet kernels \cite{seeley2014introduction} and can be viewed as low-pass filters. Rewriting $\mathbf{B}$ as follows

\begin{equation}
\mathbf{B}^{(1)}=\begin{bmatrix}\bar{\mathbf{B}}^{(1)} \\ \tilde{\mathbf{B}}^{(1)}\end{bmatrix}, \quad \mathbf{B}^{(2)}=\begin{bmatrix}\tilde{\mathbf{B}}^{(2)} \\ \bar{\mathbf{B}}^{(2)}\end{bmatrix}
\end{equation}
where $\bar{\mathbf{B}}^{(1)} \in \mathbb{C}^{(2L-1)\times (2L-1)}$,  $\tilde{\mathbf{B}}^{(1)} \in \mathbb{C}^{(2N-(2L-1)\times (2L-1)}$,$\bar{\mathbf{B}}^{(2)} \in \mathbb{C}^{(2(N-L)-1)\times (2(N-L)-1)}$,  $\tilde{\mathbf{B}}^{(2)} \in \mathbb{C}^{(2L+1)\times (2(N-L)-1)}$, 

Matrices $\tilde{\mathbf{B}}$ represent the leakage effect from out-of-band to in-band and vice versa. The primary eivenvectors of the Dirichilet kernel is known to be the global minimizer of leakage \cite{xu1984periodic}. A similar problem was addressed in \cite{10510883}, but for the continuous time kernel as opposed to the discrete sample Dirichilet kernel. Assuming that the eigenvectors are ordered in descending order according to their out-of-band leakage, the space spanned by the first $K$ eigenvectors represents the best $K$ dimensional signaling space to minimize leakage. However, the complete set of eignevectors cannot be used, since it will be a complete orthognormal set which as explained previously will have no impact on the upper bound given by \eqref{E_OCL_Ubound}. Furthermore, we must consider the other matrix factors in \eqref{leaky_x1f}. The diagonal matrix will have no impact, however, the right diagonal matrix and the up-sampling matrix must be accounted for as follows.

\begin{equation}\label{P_DPSS_spread_user1}
\mathbf{P}_{\eta}^{(1)} = \textit{D}^{-1}\left(z_{2N}^{\mathbf{k}_L(N-1)}\right)\mathbf{U}_L^T \mathcal{V}(\bar{\mathbf{B}}^{(1)})_{1:\lfloor \eta L \rfloor}
\end{equation}
\begin{equation} \label{P_DPSS_spread_user2}
\mathbf{P}_{\eta}^{(2)} = \textit{D}^{-1}\left(z_{2N}^{\mathbf{k}_{N-L}(N-1)}\right)\mathbf{U}_{N-L}^T \mathcal{V}(\bar{\mathbf{B}}^{(2)})_{1:\lfloor \eta (N-L) \rfloor}
\end{equation}

where $\mathcal{V}(\bar{\mathbf{B}}^{(1)})_{1:\lfloor \eta L \rfloor}$ is the matrix with eigenvectors of $\bar{\mathbf{B}}^{(1)}$ as its columns, $\eta<1$ represents the back-off factor which represents a reduction in the number of available signaling resources to obtain leakage reduction. 

\subsection{Spectral Efficiency for Communications vs. EISL}
In this section, we analyze the impact of spreading on communication performance in terms of capacity. For a linear time-invariant (LTI) multi-path SISO channel with additive white Gaussian noise (AWGN), the received signal in the frequency domain due to a transmission by user-1 with information symbols vector $\mathbf{s}^{(L)}\in \mathbb{C}^{L \times 1}$ across $L$ contiguous subcarriers

\begin{equation}\label{comm}
\begin{split}
\mathbf{y}^{(1)}_f &= \begin{bmatrix} \mathbf{I}_L & \mathbf{0}_{N-L} \end{bmatrix}\mathbf{F}\mathbf{H}\mathbf{F}^H\mathbf{x}_t^{(1)}+\begin{bmatrix} \mathbf{I}_L & \mathbf{0}_{N-L} \end{bmatrix}\mathbf{F}^H\mathbf{z}\\
& = D(\mathbf{h}_L) \mathbf{P}_{\eta}^{(1)} \mathbf{s}^{(1)}+\mathbf{z}_L
\end{split}
\end{equation}
where $ D(\mathbf{h}_L)$ is a diagonal matrix with subcarrier gains vector $\mathbf{h}_L$, $\mathbf{z}_L \sim \mathcal{CN}(\mathbf{0},\mathbf{I}_L)$ is AWGN.

Due to the fact that $\mathbf{P}_{\eta}^{(1)}$ is an orthogonal spreading matrix, the supported capacity is $M \lfloor  \eta L\rfloor$ where $M$ is the modulation order, and $\lfloor \eta L \rfloor$ is the number of columns in the spreading matrix.

According to the upper bound \eqref{E_OCL_Ubound}, it is possible to reduce the $E_{IB}$ by lowering the value of $\eta$. However, how the auto-correlation energy changes with $\eta$ is not known, and thus the overall impact of $\eta$ on $EISL$ is not fully known. We explore the effect of $\eta$ on the EISL in Fig. \ref{Eta_vs_EISL_M1}.

\section{Simulation Results}


\begin{figure*}[!t]
\centering
\subfloat[A-ACF for user-1 signal consisting of one OFDM symbol where $L=32$ subcarriers are allocated to user-1 and $N-L=96$ for user-2. User-2 to user-1 relative power is varied across the set of values $\alpha=-\infty , 0 , 10 ,15 , 20$ dB.]{
    \includegraphics[width=0.45\linewidth]{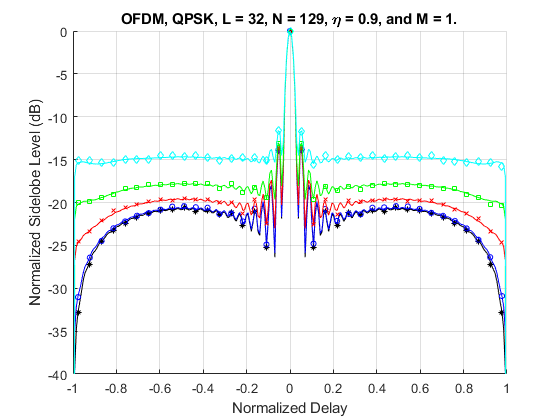}
    \label{fig:OFDM_alphas}
}
\hfill
\subfloat[A-ACF for user-1 signal consisting of 1 OFDM symbol where $L=32$ subcarriers are allocated to user-1 and $N-L=96$ for user-2, where P-DPSS spreading is applied ,$P_{0.9L}$ to user-1 and $P_{0.9(N-L)}$ to user-2. User-2 to user-1 relative power is varied across the set of values $\alpha=-\infty , 0 , 10 ,15 , 20$ dB.] {
    \includegraphics[width=0.45\linewidth]{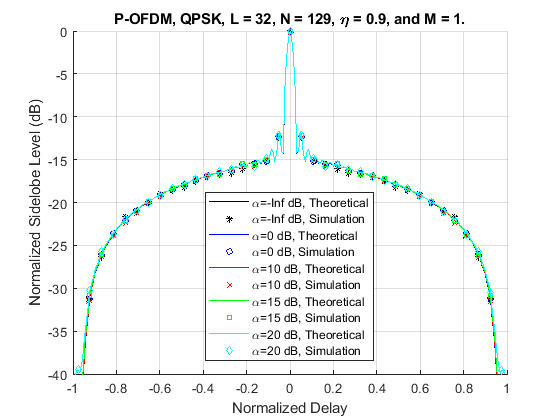}
    \label{fig:DPSS_alphas}
}

\vskip\baselineskip
\subfloat[A-ACF for user-1 signal consisting of a sequence of $M$ OFDM symbols where in each symbol $L=32$ subcarriers are allocated to user-1 and $N-L=96$ for user-2. User-2 to user-1 relative power is set to $\alpha=15 dB$]{
    \includegraphics[width=0.45\linewidth]{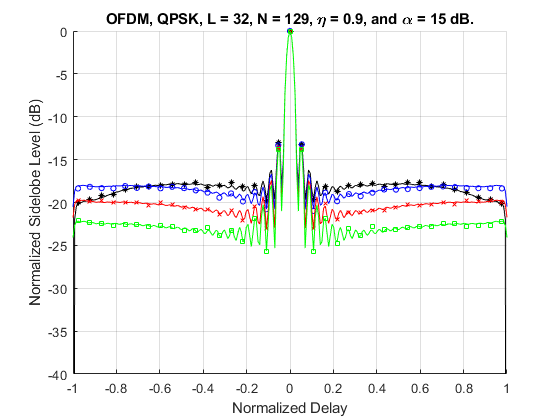}
    \label{fig:OFDM_M}
}
\hfill
\subfloat[A-ACF for user-1 signal consisting of a sequence of $M$ OFDM symbols where in each symbol $L=32$ subcarriers are allocated to user-1 and $N-L=96$ for user-2, where P-DPSS spreading is applied ,$P_{0.9L}$ to user-1 and $P_{0.9(N-L)}$ to user-2. User-2 to user-1 relative power is set to $\alpha= 15$ dB.]{
    \includegraphics[width=0.45\linewidth]{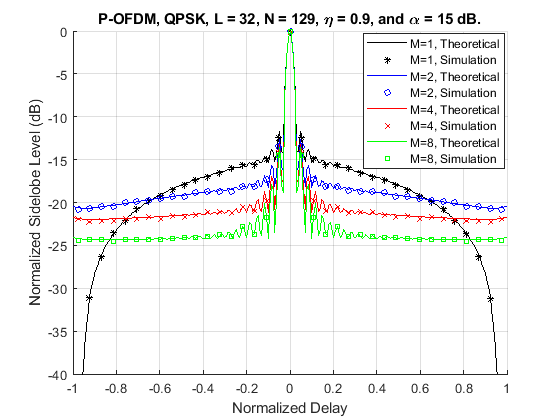}
    \label{fig:DPSS_M}
}

\caption{Comparison of OFDM and DPSS performance for varying \(\alpha\) (top row) and varying \(M\) (bottom row).}
\label{fig:A_ACF_OFDM_P_OFDM}
\end{figure*}
\begin{figure}
\centering 
\includegraphics[width=1\linewidth]{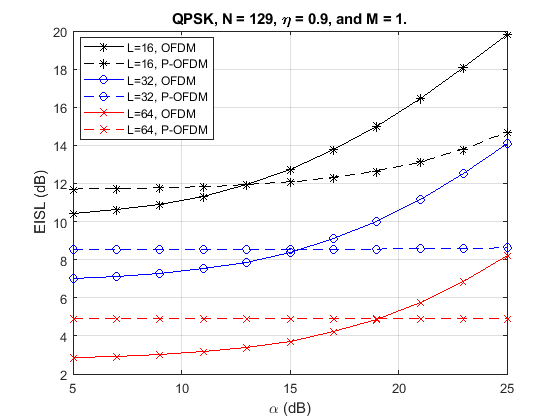}
\caption{EISL of user 1 received signal in the presence of correlation interference from user 2 vs. alpha for different numbers of subcarriers $L=16,32, 64$, using marker styles *, o, x, respectively. } \label{EISL_fig}
\end{figure}
\begin{figure}
\centering 
\includegraphics[width=1\linewidth]{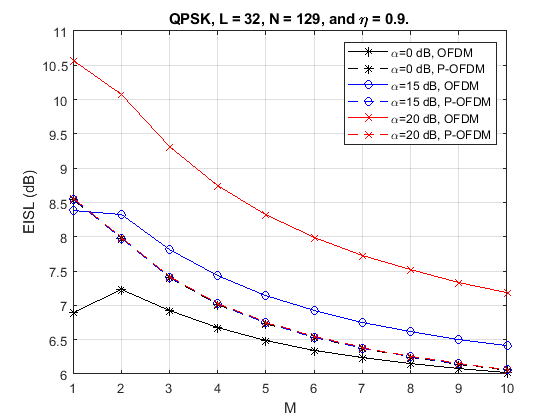}
\caption{EISL vs number of symbols in frame, $M$, for different values of relative out-of-band user power $\alpha$.} \label{EISL_vs_M_fix_alpha}
\end{figure}

\begin{figure}
\centering 
\includegraphics[width=1\linewidth]{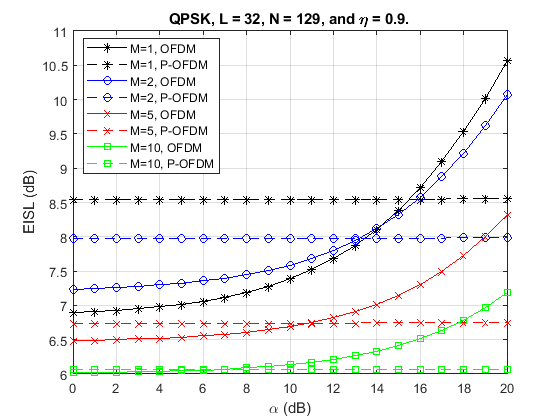}
\caption{EISL vs  relative out-of-band user power $\alpha$ for different number of symbols in a frame $M$} \label{EISL_vs_alpha}
\end{figure}


In this section, we provide simulation results for an OFDMA system consisting of two users, where user-1 does ISAC-based ranging using the back scatterings from its transmitted signal. The transmitted OFDM signal comprises a sequence of $M$ OFDM symbols, and in each symbol, $L$ subcarriers are generated from user-1, and $N-L$ sub-carriers generated from user 2 as depicted in Fig. \ref{OFDMA_ISAC}. User-1 senses an in-band back-scattered version of its transmitted signal and an out-of-band signal from user-2. The signal from user-2 is assumed to be $\alpha$ dB stronger than user-1's back-scattered signal due to the double path propagation loss.
Figure \ref{fig:A_ACF_OFDM_P_OFDM} show the A-ACF from the perspective of user-1 for an OFDM signal consisting of $M$ symbols and $L=32$ subcarriers modulated by QPSK symbols generated by user-1, and $N-L$ subcarriers modulated by QPSK symbols by user-2. The left left set of sub-figures are for OFDM with no spreading, and the right set of sub-figures involve spreading on each user acorrding to \eqref{P_DPSS_spread_user1}, \eqref{P_DPSS_spread_user2}, for user-1 and user-2, respectively, for a spectrum utilization parameter $\eta=0.9$.
To ensure a fair and practical comparison between OFDM and P-DPSS spreading, we set the spectral utilization parameter to $\eta = 0.9$ for both cases. For OFDM, this corresponds to allocating 90\% of subcarriers per user and leaving 10\% as guard bands to mitigate inter-user interference and spectral leakage. For P-DPSS, only the 90\% most spectrally concentrated basis functions are used, while the remaining 10\% are discarded due to their higher out-of-band leakage. This setup enables a realistic spectral allocation and highlights the interference suppression capability of P-DPSS under equal spectral constraints.
Each figure shows the A-ACF curve generated by simulations (dashed line) and the theoretical result  (solid line) based on \eqref{corr_vec_sq_E_matrix}. 

Figure \ref{fig:OFDM_alphas} shows A-ACF curves for a OFDM signal comprised of a single symbol, $M=1$, and different strengths of the backscattered signal according to the parameter $\alpha$ which represents the amplitude of user-2 relative to user-1. $\alpha=-\infty$ dB corresponds to the typical OFDM single user case when only a subset of $L$ subcarriers are turned on, which is depicted by the black curve and black asterisk markers. Values $\alpha = 0, 10, 15, 20$ dB correspond to curves with colors, blue, red, green, cyan, and marker styles, diamond, x, square, circle, respectively. The typical behavior of the A-ACF curve side-lobe is an oscillation around the main lobe which dies out as delay is increased, and following a concave envelope pattern with its lowest points at $\approx \pm 0.1$ and $\pm1$, and a peak at the middle point $\approx \pm 0.5$. As $\alpha$ is increased the sidelobes across the full delay range rise uniformly. For large values of $\alpha$, we can see that a $5$ dB step lead to a $5$ dB rise in the side-lobes, as predicted by the quadratic dependence on $\alpha$ in~\eqref{corr_vec_sq_E_matrix}.

Figure \ref{fig:DPSS_alphas} shows A-ACF curves for the same signal parameters in Fig. \ref{fig:OFDM_alphas} but with P-DPSS spreading applied for a spectral utilization $\eta=0.9$. We notice that P-DPSS spreading increases the side-lobes of the A-ACF compared to no spreading for small values of $\alpha$. However, the A-ACF curve stays the same across all values of $\alpha$ when spreading is applied. This shows that with P-DPSS spreading the contribution of the out-of-band cross-correlation leakage is very small such that the A-ACF stays the same as in the single user-case, i.e.,  $\alpha=-\infty$ dB.  As a result, for large values of $\alpha$, spreading will prevent the side-lobes from growing compared to OFDM. For $\alpha>20$, we can see that the A-ACF in the case of spreading is much lower compared to OFDM.

In Figures \ref{fig:OFDM_M} and \ref{fig:DPSS_M}, $\alpha $ is fixed to $15$dB and $M$ is varied over the range of values $[1,2,4,8]$. We can see that increasing $M$ lowers the side-lobes for both OFDM with and without spreading. However, we notice that for the spreading case, Fig. \ref{fig:DPSS_M}, for $M>1$, the shape of the A-ACF curve dramatically changes compared to the $M=1$ case. Thus, for $M>1$, OFDM with spreading has a more unfirom side-lobe advantage over OFDM without spreading.

for example $\alpha=15$ dB, the sie 
the amplitude difference between the in-band backscattered signal and the out-of-band signal is $20$ dB. 


\subsection{EISL Evaluation}

In Fig. \ref{EISL_fig}, EISL is plotted vs. $\alpha$ while changing the number of subcarriers allocated to user-1, $L$, and user-2, $N-L$, across the range of values $L=[16,32,64]$, depicted by black curves with '*' markers, blue curves with circle markers, red curves with 'x' markers, respectively. The solid lines are without spreading and the dashed lines are with spreading for $\eta=0.9$. We can see that larger values of $L$ EISL is improved, a reduction of $4$ to $6$ dB for every increase in $L$ by a factor of $2$. As observed previously, spread OFDM is insensitive to changes in $\alpha$, expect for large values of $\alpha $ at $L=16$ where $E_{IB}$ is comparable to the single user EISL. OFDM with spreading is better than without spreading for  $\alpha>12$ dB at $L=16$, $\alpha>15$ dB at $L=32$, and $\alpha>19$ dB at $L=64$.

Figure \ref{EISL_vs_M_fix_alpha}, shows the effect of increasing $M$ on EISL across different values of $\alpha$, $\alpha=[0,15,20]$ dB, depicted by black curves with '*' markers, blue curves with 'o' markers, red curves with 'x' markers, respectively. At $\alpha=0$ dB, for $M<10$, OFDM with no spreading is better than with spreading. At $M=10$, spreading approached the performance of the no spreading case. At $\alpha=15$ dB,  for $M>1$, spreading is better than without spreading by $\approx 0.5$ dB. At $\alpha=20$ dB, spreading is better than no spreading by $2$ dB, and the difference shrinks as $M$ is increased, reaching $1$ dB.

Figure~\ref{EISL_vs_alpha} shows the variation of EISL as a function of $\alpha$, across different values of $M$. The set of values for $M$ is $[1,2,5,10]$, depicted by black curves with '*' markers, blue curves with circle markers, red curves with `x' markers, and green curves with square markers, respectively. Solid lines represent OFDM, while dashed lines represent P-OFDM with a spectral utilization $\eta = 0.9$. For no spreading, EISL increases with $\alpha$, with increasing rates at higher values of $\alpha$ due to the domination of EISL by inter-band cross-correlation interference. For $M = 1$, the EISL curve (black solid line) shows a rapid increase beyond $\alpha = 10$ dB. Similar trends are observed for $M = 2,5,10$, but with lower initial EISL values and a slower increase with $\alpha$, indicating that increasing $M$ helps mitigate sidelobe levels. When spreading is applied, EISL remains nearly constant across all values of $\alpha$, regardless of $M$. This suggests that spreading effectively suppresses sidelobes and reduces sensitivity to inter-band cross-correlation interference. Unlike the non-spreading case, where EISL increases with $\alpha$, the curves with spreading show almost no variation across $\alpha$ for all $M$. At low $\alpha$ values and small $M$, we observe that OFDM achieves lower EISL compared to P-OFDM. This indicates that for small values of $M$, the benefit of spreading does not outweigh the increase in sidelobes caused by the spreading. However, as $M$ increases, P-OFDM consistently outperforms OFDM, showing lower EISL across most $\alpha$ values. This suggests that spreading becomes more beneficial as the number of symbols increases. At large $\alpha$ ($\alpha > 15$ dB), the EISL for the case without spreading increases sharply, whereas for the case with spreading, it remains nearly unchanged, reinforcing the advantage of spreading in mitigating interference at high power differences. Overall, Figure~\ref{EISL_vs_alpha} confirms that increasing $M$ reduces sidelobe levels, and while OFDM performs better at low $\alpha$ and small $M$, P-OFDM is more robust against power level differences and becomes superior as $M$ increases, making it advantageous in high $\alpha$ scenarios.

\begin{figure}
\centering 
\includegraphics[width=1\linewidth]{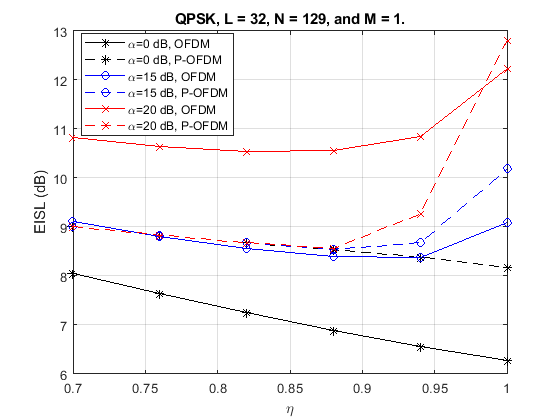}
\caption{EISL vs  spectral efficiency $\eta$ for different values of relative out-of-band user power $\alpha$ using one symbol per frame.} \label{Eta_vs_EISL_M1}
\end{figure}

Figure~\ref{Eta_vs_EISL_M1}  shows the variation of EISL with respect to $\eta$ while keeping $M = 1$ and varying $\alpha$ over $\alpha = [0, 15, 20]$ dB. Solid lines represent OFDM, while dashed lines represent P-OFDM. The black, blue, and red curves correspond to $\alpha = 0$, 15, and 20 dB, respectively. 

For no spreading, $\eta$ represents the fraction of on subcarriers, where $(1-\eta)/2$ edge subcarriers are turned off from both sides. For $\alpha=0$dB, reducing $\eta$ worsens the EISL which means that the autocorrelation side-lobes are increasing with reducing $\eta$. For high $\alpha$ values, $E_{IB}$ is more significant and as a result there is an overall reduction in EISL by  reducing $\eta$.  Reducing $\eta$ to $0.9$ results in a EISL reduction of $\approx.7$ dB at $\alpha=15 $ dB, and $\approx 1.2 $ dB at $\alpha=20 $ dB.

\begin{figure}
\centering 
\includegraphics[width=1\linewidth]{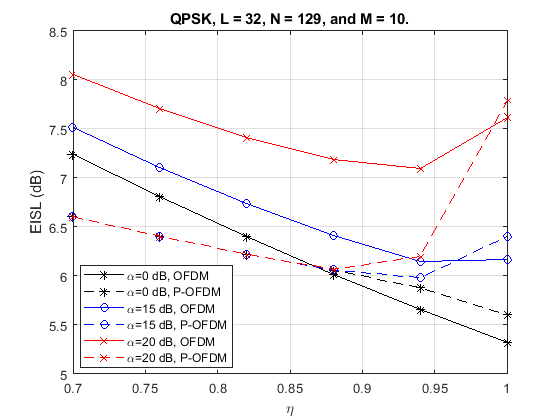}
\caption{EISL vs  spectral efficiency $\eta$ for different values of relative out-of-band user power $\alpha$ using 10 symbol per frame.} \label{Eta_vs_EISL_M10}
\end{figure}

In the case of P-DPSS spreading, for $\eta$ values lower than $0.9$, $E_{IB}$ is reduced significantly and further reduction in $\eta$ only worsens EISL due to the increase in autocorrelation side-lobes. Thus, a choice of $\eta=0.9$ for P-DPSS spreading provides the highest gain over no spreading at $\alpha = 20 $ dB. From Fig. \ref{Eta_vs_EISL_M10}, we can see that increasing $M$ to $10$ makes P-OFDM superior to OFDM starting at $\alpha=15 $dB. Further improvements are expected for higher values of $M$ as indicated in Figures \ref{EISL_vs_alpha} and \ref{EISL_vs_M_fix_alpha}.

\begin{figure}
\centering 
\includegraphics[width=1\linewidth]{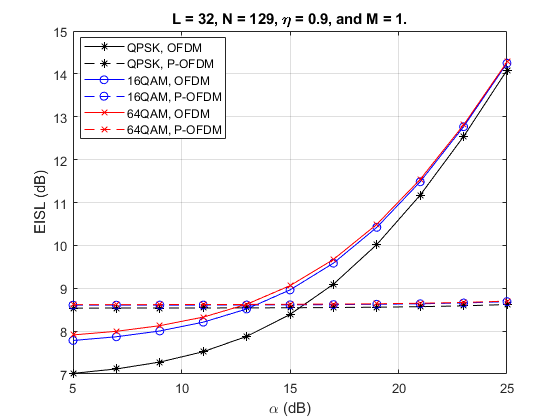}
\caption{EISL vs relative out-of-band user power $\alpha$ for different modulation schemes.} \label{EISL_vs_Alpha_vs_mod}
\end{figure}

Figure~\ref{EISL_vs_Alpha_vs_mod} shows EISL versus $\alpha$ for QPSK, 16QAM, and 64QAM using OFDM (solid lines) and P-OFDM (dashed lines). For OFDM, EISL increases with $\alpha$, and higher-order modulations yield higher EISL. In contrast, P-OFDM maintains nearly flat EISL across all $\alpha$ and modulation formats. At low $\alpha$, OFDM slightly outperforms P-OFDM, especially with QPSK, but P-OFDM becomes superior at high $\alpha$, highlighting its robustness to power imbalance and modulation order.

\section{Conclusion}
In this paper, we investigated the impact of out-of-band correlation leakage in multi-user OFDMA-based ISAC systems. Our analysis demonstrated that while OFDM is globally optimal for minimizing periodic sidelobe levels, its performance degrades in terms of aperiodic correlation, especially when adjacent frequency interference is present. We characterized this out-of-band correlation leakage energy $E_{IB}$ and showed that it can significantly increase the integrated sidelobe levels, quantified by the $EISL$ metric. Minimizing $EISL$ is crucial for ranging accuracy in the presence of multiple targets. To address this, we proposed a spreading scheme that transforms the OFDM basis to a DPSS basis, thereby minimizing OBCL and achieving better sidelobe suppression. Simulation results validate our claims, showing that the proposed scheme effectively reduces ISL in multi-user ISAC setups. Future work will focus on optimizing the spreading scheme parameters and exploring other orthonormal bases to further enhance ISAC performance in diverse multi-user environments.

\begin{appendices}
\section{ }\label{apndx_A}
Starting from the second line of \eqref{leaky_x1f} and ignoring the spreading matrix $\mathbf{P}^{(1)}$, we expand \(\bar{\mathbf{x}}^{(1)}_{f}[k]\):

\begin{equation}\label{up_sampleby2_in_f}
\begin{split}
\bar{\mathbf{x}}^{(1)}_{f}[k] &= \sum_{n=0}^{N-1} e^{-j \frac{2\pi k n}{2N}} \left( \sum_{k'=0}^{L-1} e^{j \frac{2 \pi n k'}{N}} \mathbf{s}[k'] \right)\\
&= z^{-k\frac{N-1}{2N}}\sum_{k' = 0}^{L-1} \frac{\sin((0.5k - k')\pi)}{\sin((0.5k - k')\frac{\pi}{N})} z^{k'\frac{N-1}{2N}} \mathbf{s}[k'] \\
&= z^{-k\frac{N-1}{2N}} \sum_{\substack{k' = 0, \\ \text{even}}}^{2L-2} \frac{\sin(0.5(k - k')\pi)}{\sin(0.5(k - k')\frac{\pi}{N})} z^{k'\frac{N-1}{N}}  \mathbf{s}[0.5k'] \\
&= z^{-k\frac{N-1}{2N}} \sum_{k'=0}^{2L-1} \frac{\sin(N(k - k')\frac{\pi}{2N})}{\sin((k - k')\frac{\pi}{2N})} (z^{k'\frac{N-1}{2N}}  \mathbf{s}[k'])\uparrow^2 \\
\end{split}
\end{equation}
where $\uparrow^2$ denotes up-sampling by 2.

We can rewrite the last line of \eqref{up_sampleby2_in_f} in matrix form
\begin{equation}\label{up_sampleBy2_f_x1}
\begin{split}
\bar{\mathbf{x}}^{(1)}_f&=\textit{D}\left(\mathbf{z}_{2N}^{\frac{N-1}{2N}}\right) \mathbf{B}^{(1)}\mathbf{U}_L\textit{D}\left(\mathbf{z}_{L}^{\frac{N-1}{2N}}\right)\mathbf{s}^{(1)}\\
\end{split}
\end{equation}

Similarly, for \(\bar{\mathbf{x}}_{2f}[k]\):
\begin{equation}\label{up_sampleBy2_f_x1}
\begin{split}
\bar{\mathbf{x}}^{(2)}_f&=\textit{D}\left(\mathbf{z}_{2N}^{\frac{N-1}{2N}}\right) \mathbf{B}^{(2)}\mathbf{U}_{N-L}\textit{D}\left(\mathbf{z}_{N-L}^{\frac{N-1}{2N}}\right)\mathbf{s}^{(2)}\\
\end{split}
\end{equation}
where $\mathbf{U}_L=[\mathbf{e}_1,\mathbf{0}, \mathbf{e}_2,..,\mathbf{0},\mathbf{e}_L]^T$ is an up-sampling by 2 matrix, where $[\mathbf{e}_i]_k=\delta(k-i), k =0,..,L-1 $. 
$[\mathbf{B}]_{k,k'}=\frac{\sin(N(k - k')\frac{\pi}{2N})}{\sin((k - k')\frac{\pi}{2N})}$   $ \text{ for }k,k'=0,..,2N-1$.
\begin{equation}
\begin{split}
\mathbf{B}&= \begin{bmatrix}
\mathbf{B}^{(1)} & \mathbf{b}_{2L-1} &\mathbf{B}^{(2)} & \mathbf{b}_{2N-1} \\
\end{bmatrix}\\
\end{split}
\end{equation}

\begin{equation}
\mathbf{B}^{(1)} = [\bar{\mathbf{B}}^{(1)T},\tilde{\mathbf{B}}^{(1)T}]^T, \mathbf{B}^{(2)} = [\tilde{\mathbf{B}}^{(2)T},\bar{\mathbf{B}}^{(2)T}]^T
\end{equation}
where $\mathbf{b}_r=\frac{\sin(N(k - r)\frac{\pi}{2N})}{\sin((k - r)\frac{\pi}{2N})}$   $ \text{ for }k=0,..,2N-1$,
 $\bar{\mathbf{B}}^{(1)}\in \mathbb{C}^{(2L-1)\times (2 L-1)}$, $\tilde{\mathbf{B}}^{(1)}\in \mathbb{C}^{(2L'+1) \times (2L-1) }$ ,$\bar{\mathbf{B}}^{(2)}\in \mathbb{C}^{(2L'-1)\times (2 L'-1)}$, $\tilde{\mathbf{B}}^{(2)}\in \mathbb{C}^{(2L+1)\times (2L'-1)}$.
 $L'=N-L$.

\section{Proof of Proposition 1}\label{apndx_B}

Substituting in \eqref{mult_sym_corr} using  $\bar{\mathbf{x}}^{(1)}_f=\mathbf{V}^{(1)H}\mathbf{s}^{(1)}, \bar{\mathbf{x}}^{(2)}_f=\mathbf{V}^{(2)H}\mathbf{s}^{(2)} $


\begin{equation}\label{mult_sym_corr_2}
\begin{split}
&\mathbf{c}_{\mathbf{x}_{\mathbf{M}},\mathbf{y}_{\mathbf{M}}}[k]\\
&=\sum_{i,n}\left(\mathbf{v}_n^{(1)H}\mathbf{s}^{(1)}_{i}\mathbf{v}^{(1)T}_n \mathbf{s}_{i}^{(1)*}+\alpha\mathbf{v}_n^{(1)H}\mathbf{s}^{(1)}_{i}\mathbf{v}_n^{(2)T}\mathbf{s}_{i}^{(1)*}\right)z^{\frac{mk}{N}}\\
&+\sum_{j,m}\left(\mathbf{v}_{m}^{(1)H}\mathbf{s}^{(1)}_{j}\mathbf{v}^{(1)T}_{m}\mathbf{s}_{j+1}^{(1)*}+\alpha\mathbf{v}_{m}^{(1)H}\mathbf{s}^{(1)}_{j}\mathbf{v}_{m}^{(2)T}\mathbf{s}_{j+1}^{(2)*}\right)\frac{z^{\frac{mk}{N}}}{e^{j\pi m}}\\
&=\sum_{n}(\mathbf{v}_{n}^{(1)T}\otimes\mathbf{v}_{n}^{(1)H})z^{\frac{kn}{N}}\sum_i(\mathbf{s}^{(1)*}_{i}\otimes\mathbf{s}^{(1)}_{i})\\
&+\alpha\sum_n(\mathbf{v}_{n}^{(2)T}\otimes\mathbf{v}_{n}^{(1)H})z^{\frac{kn}{N}}\sum_i(\mathbf{s}_{i}^{(2)*}\otimes\mathbf{s}^{(1)}_{i})\\
&+\sum_m (\mathbf{v}_{m}^{(1)T}\otimes\mathbf{v}_{m}^{(1)H})(-z^{\frac{k}{N}})^m\sum_j(\mathbf{s}^{(1)*}_{j+1}\otimes\mathbf{s}^{(1)}_{j})+\\
&\alpha \sum_m(\mathbf{v}_{m}^{(2)T}\otimes\mathbf{v}_{m}^{(1)H})(-z^{\frac{k}{N}})^m\sum_j(\mathbf{s}^{(2)*}_{j+1}\otimes\mathbf{s}^{(1)}_{j})\\
\end{split}
\end{equation}
the upper limits of the sums are removed for the sake of compactness since they can be known for their corresponding variables, where $i=0,..,M-1, j=0,..,M-2, n,m=0,..,2N-1$ .

Let
\begin{equation}\label{Sigma_v}
\mathbf{\Sigma}_{\mathbf{v}}^{(l,k)}=\sum_{n}(\mathbf{v}_{n}^{(l)T}\otimes\mathbf{v}_{n}^{(k)H})z^{\frac{kn}{N}}
\end{equation}
\begin{equation}\label{tilde_Sigma_v}
\tilde{\mathbf{\Sigma}}_{\mathbf{v}}^{(l,k)}=\sum_{n}(\mathbf{v}_{n}^{(l)T}\otimes\mathbf{v}_{n}^{(k)H})(-z^{\frac{k}{N}})^n
\end{equation}
Substituting into \eqref{mult_sym_corr_2} and doing further manipulation
\begin{equation}\label{mult_sym_corr_2}
\begin{split}
&\mathbf{c}_{\mathbf{x}_{\mathbf{M}},\mathbf{y}_{\mathbf{M}}}[k]\\
&=\mathbf{\Sigma}_{\mathbf{v}}^{(1,1)}\sum_i(\mathbf{s}_{i}^{(1)*}\otimes\mathbf{s}^{(1)}_{i})+\alpha\mathbf{\Sigma}_{\mathbf{v}}^{(2,1)}\sum_i(\mathbf{s}_{i}^{(2)*}\otimes\mathbf{s}_{i}^{(1)})\\
&+\tilde{\mathbf{\Sigma}}_{\mathbf{v}}^{(1,1)}\sum_j(\mathbf{s}_{j+1}^{(1)*}\otimes\mathbf{s}_{j}^{(1)})+\alpha \tilde{\mathbf{\Sigma}}_{\mathbf{v}}^{(2,1)}\sum_j(\mathbf{s}_{j+1}^{(2)*}\otimes\mathbf{s}_{j}^{(1)})\\
\end{split}
\end{equation}

Now we find $\mathbb{E}(|\mathbf{c}_{\mathbf{x}_{\mathbf{M}},\mathbf{y}_M}[k]|^2)$. 
By the independence of symbol vectors $\mathbf{s}^{(1)}$, $\mathbf{s}^{(2)}$ 

\begin{equation}\label{E_CxMyM}
\begin{split}
&\mathbb{E}\left(|\mathbf{\Sigma}_{\mathbf{v}}^{(1,1)}\sum_i(\mathbf{s}_{i}^{(1)*}\otimes\mathbf{s}_{i}^{(1)})
+\tilde{\mathbf{\Sigma}}_{\mathbf{v}}^{(1,1)}\sum_j(\mathbf{s}_{j+1}^{(1)*}\otimes\mathbf{s}_{j}^{(1)})|^2\right)\\
&+\alpha^2\mathbb{E}\left(|\mathbf{\Sigma}_{\mathbf{v}}^{(2,1)}\sum_i(\mathbf{s}_{i}^{(2)*}\otimes\mathbf{s}_{i}^{(1)})+ \tilde{\mathbf{\Sigma}}_{\mathbf{v}}^{(2,1)}\sum_j(\mathbf{s}_{j+1}^{(2)*}\otimes\mathbf{s}_{j}^{(1)})|^2\right)\\
\end{split}
\end{equation}

The first term in \eqref{E_CxMyM} is simplified as follows
\begin{equation*}\label{E_CxMyM_1st}
\begin{split}
&\mathbb{E}\left(|\mathbf{\Sigma}_{\mathbf{v}}^{(1,1)}\sum_i(\mathbf{s}_{i}^{(1)*}\otimes\mathbf{s}_{i}^{(1)})
+\tilde{\mathbf{\Sigma}}_{\mathbf{v}}^{(1,1)}\sum_j(\mathbf{s}_{j+1}^{(1)*}\otimes\mathbf{s}_{j}^{(1)})|^2\right)\\
&=\mathbf{\Sigma}_{\mathbf{v}}^{(1,1)}\sum_{i,i'}\mathbb{E}((\mathbf{s}_{i}^{(1)*}\otimes\mathbf{s}_{i}^{(1)})(\mathbf{s}_{i'}^{(1)T}\otimes\mathbf{s}_{i'}^{(1)H}))\mathbf{\Sigma}_{\mathbf{v}}^{(1,1)H}\\
&+2\Re\left(\mathbf{\Sigma}_{\mathbf{v}}^{(1,1)}\sum_{i,j}\mathbb{E}((\mathbf{s}_{i}^{(1)*}\otimes\mathbf{s}_{i}^{(1)})(\mathbf{s}_{j+1}^{(1)T}\otimes\mathbf{s}_{j}^{(1)H}))\tilde{\mathbf{\Sigma}}_{\mathbf{v}}^{(1,1)H}\right)\\
&+\tilde{\mathbf{\Sigma}}_{\mathbf{v}}^{(1,1)}\sum_{j,j'}\mathbb{E}(\mathbf{s}_{j+1}^{(1)*}\mathbf{s}_{j'+1}^{(1)T}\otimes\mathbf{s}_{j}^{(1)}\mathbf{s}_{j'}^{(1)H})\tilde{\mathbf{\Sigma}}_{\mathbf{v}}^{(1,1)H}\\
&=\mathbf{\Sigma}_{\mathbf{v}}^{(1,1)}\mathbf{S}\mathbf{\Sigma}_{\mathbf{v}}^{(1,1)H}\\
&+\mathbf{\Sigma}_{\mathbf{v}}^{(1,1)}\sum_{i\neq i'}\mathbb{E}((\mathbf{s}_{i}^{(1)*}\otimes\mathbf{s}_{i}^{(1)})(\mathbf{s}_{i'}^{(1)T}\otimes\mathbf{s}_{i'}^{(1)H}))\mathbf{\Sigma}_{\mathbf{v}}^{(1,1)H}\\
&+\tilde{\mathbf{\Sigma}}_{\mathbf{v}}^{(1,1)}\sum_{j}\mathbb{E}(\mathbf{s}_{j+1}^{(1)*}\mathbf{s}_{j+1}^{(1)T})\otimes\mathbb{E}(\mathbf{s}_{j}^{(1)}\mathbf{s}_{j}^{(1)H})\tilde{\mathbf{\Sigma}}_{\mathbf{v}}^{(1,1)H}\\
&=M\mathbf{\Sigma}_{\mathbf{v}}^{(1,1)}\mathbf{S}\mathbf{\Sigma}_{\mathbf{v}}^{(1,1)H}\\
&+\mathbf{\Sigma}_{\mathbf{v}}^{(1,1)}\sum_{i\neq i'}\mathbb{E}(\mathbf{s}_{i}^{(1)*}\otimes\mathbf{s}_{i}^{(1)})\mathbb{E}(\mathbf{s}_{i'}^{(1)T}\otimes\mathbf{s}_{i'}^{(1)H})\mathbf{\Sigma}_{\mathbf{v}}^{(1,1)H}\\
&+\tilde{\mathbf{\Sigma}}_{\mathbf{v}}^{(1,1)}\sum_{j}\mathbb{E}(\mathbf{s}_{j+1}^{(1)*}\mathbf{s}_{j+1}^{(1)T})\otimes\mathbb{E}(\mathbf{s}_{j}^{(1)}\mathbf{s}_{j}^{(1)H})\tilde{\mathbf{\Sigma}}_{\mathbf{v}}^{(1,1)H}\\
&=M\mathbf{\Sigma}_{\mathbf{v}}^{(1,1)}\mathbf{S}\mathbf{\Sigma}_{\mathbf{v}}^{(1,1)H}\\
&+\mathbf{\Sigma}_{\mathbf{v}}^{(1,1)}\sum_{i\neq i'}\mathbb{E}(\mathbf{s}_{i}^{(1)*}\otimes\mathbf{s}_{i}^{(1)})\mathbb{E}(\mathbf{s}_{i'}^{(1)T}\otimes\mathbf{s}_{i'}^{(1)H})\mathbf{\Sigma}_{\mathbf{v}}^{(1,1)H}\\
&+\tilde{\mathbf{\Sigma}}_{\mathbf{v}}^{(1,1)}\sum_{j}(\mathbf{I}_L\otimes\mathbf{I}_L)\tilde{\mathbf{\Sigma}}_{\mathbf{v}}^{(1,1)H}\\
\end{split}
\end{equation*}
\begin{equation}\label{E_CxMyM_1st_2nd_part}
\begin{split}
&=M\mathbf{\Sigma}_{\mathbf{v}}^{(1,1)}\mathbf{S}\mathbf{\Sigma}_{\mathbf{v}}^{(1,1)H}\\
&+\mathbf{\Sigma}_{\mathbf{v}}^{(1,1)}(M^2-M)\mathbf{S}_2\mathbf{\Sigma}_{\mathbf{v}}^{(1,1)H}+\tilde{\mathbf{\Sigma}}_{\mathbf{v}}^{(1,1)}\sum_{j}\mathbf{I}_{L^2}\tilde{\mathbf{\Sigma}}_{\mathbf{v}}^{(1,1)H}\\
&=M\mathbf{\Sigma}_{\mathbf{v}}^{(1,1)}\mathbf{S}\mathbf{\Sigma}_{\mathbf{v}}^{(1,1)H}+(M^2-M)\mathbf{\Sigma}_{\mathbf{v}}^{(1,1)}\mathbf{S}_2\mathbf{\Sigma}_{\mathbf{v}}^{(1,1)H}\\
&+(M-1)\tilde{\mathbf{\Sigma}}_{\mathbf{v}}^{(1,1)}\tilde{\mathbf{\Sigma}}_{\mathbf{v}}^{(1,1)H}\\
\end{split}
\end{equation}
where $\mathbf{S}\triangleq \mathbb{E}(\mathbf{s}^*\mathbf{s}^T\otimes\mathbf{s}\mathbf{s}^H )$ and can be decomposed as in \eqref{S_kurtosis}.

\begin{equation}\label{S_kurtosis}
\mathbf{S}=\mathbf{I}_{L^2} + \mathbf{S}_1+\mathbf{S}_2
\end{equation}
where
$\mathbf{S}_1=\text{diag}\left([\mu_4-2,\mathbf{0}_L^T,\mu_4-2,\mathbf{0}^T_L,...,\mu_4-2]^T\right)$
$\mathbf{S}_2=[\mathbf{c},\mathbf{0}_{L^2\times L},\mathbf{c},...,\mathbf{c},\mathbf{0}_{L^2\times L},\mathbf{c}]$
$\mathbf{c}=[1,\mathbf{0}_L^T,1,...1,\mathbf{0}_L^T,1]^T$ and the derivation can be found in \cite{liu2024ofdm}.

Substituting using \eqref{Sigma_v} and \eqref{tilde_Sigma_v} into the last line of \eqref{E_CxMyM_1st_2nd_part}, the expression simplifies to
 \begin{equation}\label{E_CxMyM_1st_smplfd}
\begin{split}
&M\mathbb{E}(|\mathbf{c}_{11}[k]|^2)+(M^2-M)\sum_{m,n}||\mathbf{v}^{(1)}_n||^2 ||\mathbf{v}^{(1)}_m||^2z^{\frac{k(n-m)}{N}}\\
&+(M-1)\sum_{m,n}|\mathbf{v}^{(1)H}_n\mathbf{v}^{(1)}_m|^2(-z^{\frac{k}{N}})^{n-m}\\
\end{split}
\end{equation}

The term $\mathbb{E}(|\mathbf{c}_{11}[k]|^2)$ in \eqref{E_CxMyM_2nd} is given by \eqref{x1_autocorr}, which represents the A-ACF in the single user case,.

\begin{equation}\label{x1_autocorr}
\begin{split}
&\mathbb{E}(|\mathbf{c}_{11}[k]|^2)\\
& = \sum_{m,n=0}^{2N-1}z_k^{n-m} \biggl[  |\mathbf{v}_n^{(1)H}\mathbf{v}^{(1)}_m|^2+(\mu_4-2)||\mathbf{v}^{(1)}_n \odot \mathbf{v}^{(1)}_m||_2^2\\
&+||\mathbf{v}^{(1)}_n||_2^2||\mathbf{v}^{(1)}_m||_2^2 \biggr]
\end{split}
\end{equation}
The second term in \eqref{E_CxMyM} simplifies to
\begin{equation}\label{E_CxMyM_2nd}
\begin{split}
&\alpha^2\mathbf{\Sigma}_{\mathbf{v}}^{(2,1)}\sum_{i,i'}\mathbb{E}(\mathbf{s}_{2,i}^*\mathbf{s}_{2,i'}^T\otimes\mathbf{s}_{1,i}\mathbf{s}_{1,i'}^H)\mathbf{\Sigma}_{\mathbf{v}}^{(2,1)H}\\
&+\alpha^2\mathbf{\Sigma}_{\mathbf{v}}^{(2,1)}\sum_{i,j}\mathbb{E}(\mathbf{s}_{2,i}^*\mathbf{s}_{2,j+1}^T\otimes\mathbf{s}_{1,i}\mathbf{s}_{1,j}^H)\tilde{\mathbf{\Sigma}}_{\mathbf{v}}^{(2,1)H}\\
&+\alpha^2\tilde{\mathbf{\Sigma}
}_{\mathbf{v}}^{(2,1)}\sum_{j,i}\mathbb{E}(\mathbf{s}_{2,j+1}^*\mathbf{s}_{2,i}^T\otimes\mathbf{s}_{1,j}\mathbf{s}_{1,i}^H)\mathbf{\Sigma}_{\mathbf{v}}^{(2,1)H}\\
&+\alpha^2\tilde{\mathbf{\Sigma}}_{\mathbf{v}}^{(2,1)}\sum_{j,j'}\mathbb{E}(\mathbf{s}_{2,j+1}^*\mathbf{s}_{2,j'+1}^T\otimes\mathbf{s}_{1,j}\mathbf{s}_{1,j'}^H)\tilde{\mathbf{\Sigma}}_{\mathbf{v}}^{(2,1)H}\\
&=\alpha^2\mathbf{\Sigma}_{\mathbf{v}}^{(2,1)}\sum_{i}\mathbb{E}(\mathbf{s}_{2,i}^*\mathbf{s}_{2,i}^T)\otimes\mathbb{E}(\mathbf{s}_{1,i}\mathbf{s}_{1,i}^H)\mathbf{\Sigma}_{\mathbf{v}}^{(2,1)H}\\
&+\alpha^2\tilde{\mathbf{\Sigma}}_{\mathbf{v}}^{(2,1)}\sum_{j}\mathbb{E}(\mathbf{s}_{2,j+1}^*\mathbf{s}_{2,j+1}^T)\otimes\mathbb{E}(\mathbf{s}_{1,j}\mathbf{s}_{1,j}^H)\tilde{\mathbf{\Sigma}}_{\mathbf{v}}^{(2,1)H}\\
&=\alpha^2M\mathbf{\Sigma}_{\mathbf{v}}^{(2,1)}\mathbf{\Sigma}_{\mathbf{v}}^{(2,1)H}+\alpha^2(M-1)\tilde{\mathbf{\Sigma}}_{\mathbf{v}}^{(2,1)}\tilde{\mathbf{\Sigma}}_{\mathbf{v}}^{(2,1)H}\\
&=\alpha^2M\sum_{m,n}\mathbf{v}^{(2)T}_n\mathbf{v}^{(2)*}_m\mathbf{v}^{(1)H}_n\mathbf{v}^{(1)}_mz_k^{n-m}\\
&+\alpha^2(M-1)\sum_{m,n}\mathbf{v}^{(2)T}_n\mathbf{v}^{(2)*}_m\mathbf{v}^{(1)H}_n\mathbf{v}^{(1)}_m(-z_k)^{n-m}
\end{split}
\end{equation}

Summing \eqref{E_CxMyM_1st_smplfd} and the last line in \eqref{E_CxMyM_2nd}, and normalizing w.r.t. the main lobe energy $(ML)^2$  , we get \eqref{corr_vec_sq_E}.

\section{Proof of Corollary 1}\label{apndx_C}
 
\begin{equation}\label{EISL}
\begin{split}
&EISL\\
&= \sum_{k=(-N-1)}^{N-1}\mathbb{E}(|\mathbf{c}_{\mathbf{x}_{\mathbf{M}}\mathbf{y}_{\mathbf{M}}}[k]|^2)-\mathbb{E}(|\mathbf{c}_{\mathbf{x}_{\mathbf{M}}\mathbf{y}_{\mathbf{M}}}[0]|^2)\\
&=  2N\sum_{n=0}^{2N-1} \left(  \frac{M^2+2M-1}{(ML)^2}||\mathbf{v}_n||_2^4+\frac{M}{(ML)^2}(\mu_4-2)||\mathbf{v}_n||_4^4\right)\\
&+2N\alpha^2\sum_{k=0}^{2N-1}\frac{2M-1}{(ML)^2}||\mathbf{v}_n||_2^2||\mathbf{w}_n||_2^2- \frac{\mu_4 -1+ML}{(M*L)^2}\\
\end{split}
\end{equation}

In the second line of \eqref{EISL} we used the following fact 
$\mathbb{E}(|\mathbf{c}_{\mathbf{x}_{\mathbf{M}}\mathbf{y}_{\mathbf{M}}}[0]|^2)= \mathbb{E}(|\mathbf{c}_{\mathbf{x}_{\mathbf{M}}\mathbf{x}_{\mathbf{M}}}[0]|^2)= \mathbb{E}(\sum_{m=0}^{M-1}||\mathbf{s}_m||_2^2)
= (\mu_4 -1+ML)ML$

\section{Proof of Theorem}\label{apndx_D}

Let $\mathbf{W}^{(1)}=\mathbf{V}^{(1)H}$ and $\mathbf{W}^{(2)}=\mathbf{V}^{(2)H}$ where

\begin{equation}\label{W1_W2_submatrices}
\mathbf{W}^{(1)}=[
\bar{\mathbf{W}}^{(1)H}\\
\tilde{\mathbf{W}}^{(1)H}
]^H,\quad \mathbf{W}^{(2)} = 
[\tilde{\mathbf{W}}^{(2)H}\\
\bar{\mathbf{W}}^{(2)H}]^H
\end{equation}
where $\bar{\mathbf{W}}^{(1)} \in \mathbb{C}^{2L\times L}$, $\tilde{\mathbf{W}}^{(1)} \in \mathbb{C}^{2(N-L)\times L}$, $\tilde{\mathbf{W}}^{(2)} \in \mathbb{C}^{2L\times (N-L)}$,  $\bar{\mathbf{W}}^{(2)} \in \mathbb{C}^{2(N-L)\times (N-L)}$.

Substituting \eqref{W1_W2_submatrices} into \eqref{E_OCL}

\begin{equation} \label{E_OCL_mat_W}
\begin{split}
E_{IB}&=
\text{tr}(\bar{\mathbf{W}}^{(1)}\bar{\mathbf{W}}^{(1)H} \odot \tilde{\mathbf{W}}^{(2)}\tilde{\mathbf{W}}^{(2)H})\\
&+ \text{tr}(\tilde{\mathbf{W}}^{(1)}\tilde{\mathbf{W}}^{(1)H }\odot \bar{\mathbf{W}}^{(2)}\bar{\mathbf{W}}^{(2)H})\\
& =\text{tr}( \sum_{i,j=0}^{L-1}\text{diag}(\bar{\mathbf{w}}_i^{(1)}\bar{\mathbf{w}}_j^{(1)H} )\sum_{i',j'=0}^{N-L-1} \text{diag(}\tilde{\mathbf{w}}_{i'}^{(2)}\tilde{\mathbf{w}}_{j'}^{(2)H}))\\
&+\text{tr}( \sum_{i,j=0}^{L-1}\text{diag}(\tilde{\mathbf{w}}_i^{(1)}\tilde{\mathbf{w}}_j^{(1)H} )\sum_{i',j'=0}^{N-L-1} \text{diag}(\bar{\mathbf{w}}_{i'}^{(2)}\bar{\mathbf{w}}_{j'}^{(2)H}))\\
& =( \sum_{i,j=0}^{L-1}\bar{\mathbf{w}}_i^{(1)}\odot\bar{\mathbf{w}}_j^{(1)*} )^H(\sum_{i',j'=0}^{N-L-1} \tilde{\mathbf{w}}_{i'}^{(2)}\odot\tilde{\mathbf{w}}_{j'}^{(2)*})\\
&+( \sum_{i,j=0}^{L-1}\tilde{\mathbf{w}}_i^{(1)}\odot\tilde{\mathbf{w}}_j^{(1)*})^H( \sum_{i',j'=0}^{N-L-1} \bar{\mathbf{w}}_{i'}^{(2)}\odot\bar{\mathbf{w}}_{j'}^{(2)*})\\
\end{split}
\end{equation}

Using the Cauchy-Schwarz inequality, we obtain the following upper-bound
\begin{equation*}
\begin{split}
E_{IB}&\leq \sqrt{||\sum_{i,j=0}^{L-1}\bar{\mathbf{w}}_i^{(1)}\odot\bar{\mathbf{w}}_j^{(1)*}||_2^2||\sum_{i',j'=0}^{N-L-1} \tilde{\mathbf{w}}_{i'}^{(2)}\odot\tilde{\mathbf{w}}_{j'}^{(2)*}||_2^2}\\
&+ \sqrt{||\sum_{i,j=0}^{L-1}\tilde{\mathbf{w}}_i^{(1)}\odot\tilde{\mathbf{w}}_j^{(1)*}||_2^2||\sum_{i',j'=0}^{N-L-1} \bar{\mathbf{w}}_{i'}^{(2)}\odot\bar{\mathbf{w}}_{j'}^{(2)*}||_2^2}\\
\end{split}
\end{equation*}
\begin{equation}
\begin{split}
&\leq \sqrt{\sum_{i,j=0}^{L-1}||\bar{\mathbf{w}}_i^{(1)}\odot\bar{\mathbf{w}}_j^{(1)*}||_2^2\sum_{i',j'=0}^{N-L-1} ||\tilde{\mathbf{w}}_{i'}^{(2)}\odot\tilde{\mathbf{w}}_{j'}^{(2)*}||_2^2}\\
&+ \sqrt{\sum_{i,j=0}^{L-1}||\tilde{\mathbf{w}}_i^{(1)}\odot\tilde{\mathbf{w}}_j^{(1)*}||_2^2\sum_{i',j'=0}^{N-L-1} ||\bar{\mathbf{w}}_{i'}^{(2)}\odot\bar{\mathbf{w}}_{j'}^{(2)*}||_2^2}\\
&\leq \sqrt{\sum_{i,j=0}^{L-1}||\bar{\mathbf{w}}_i^{(1)}||_2^2||\bar{\mathbf{w}}_j^{(1)*}||_2^2\sum_{i',j'=0}^{N-L-1} ||\tilde{\mathbf{w}}_{i'}^{(2)}||_2^2||\tilde{\mathbf{w}}_{j'}^{(2)*}||_2^2}\\
&+ \sqrt{\sum_{i,j=0}^{L-1}||\tilde{\mathbf{w}}_i^{(1)}||_2^2||_2^2||\tilde{\mathbf{w}}_j^{(1)*}||_2^2\sum_{i',j'=0}^{N-L-1} ||\bar{\mathbf{w}}_{i'}^{(2)}||_2^2||\bar{\mathbf{w}}_{j'}^{(2)*}||_2^2}\\
&= \sqrt{||\bar{\mathbf{W}}^{(1)}||^4_F ||\tilde{\mathbf{W}}^{(2)}||^4_F} + \sqrt{||\tilde{\mathbf{W}}^{(1)}||^4_F ||\bar{\mathbf{W}}^{(2)}||^4_F} \\
&= ||\bar{\mathbf{W}}^{(1)}||^2_F ||\tilde{\mathbf{W}}^{(2)}||^2_F + ||\tilde{\mathbf{W}}^{(1)}||^2_F ||\bar{\mathbf{W}}^{(2)}||^2_F 
\end{split}
\end{equation}
\end{appendices}
\IEEEpeerreviewmaketitle


%

\bibliographystyle{IEEEtran}
\bibliography{IEEEabrv,references.bib}

\end{document}